\documentclass[12pt]{iopart}
\usepackage{graphicx}  % needed for figures
\usepackage{dcolumn}   % needed for some tables
\usepackage[usenames]{color}
\usepackage{iopams}  
\usepackage{psfrag}
\begin{document}

\newcommand{\bra}[1]{\mbox{\ensuremath{\langle #1 \vert}}}
\newcommand{\ket}[1]{\mbox{\ensuremath{\vert #1 \rangle}}}
\newcommand{\lket}[2]{\tiny{$|\!(\! #1 \!)\!,\!N\!_0\!+\! #2\rangle$}}
\newcommand{\lketm}[2]{\tiny{$|\!(\! #1 \!)\!,\!N\!_0\!-\! #2\!\rangle$}}
\newcommand{\lketz}[1]{\tiny{$|\!(\! #1 \!)\!,\!N\!_0\!\rangle$}}
\newcommand{\mb}[1]{\mathbf{#1}}
\newcommand{\rbs}{$^{87}$Rb}
\newcommand{\kf}{$^{40}$K}
\newcommand{\na}{${^{23}}$Na}
\newcommand{\lisix}{$^{6}$Li}
\newcommand{\muK}{\:\mu\textrm{K}}
\newcommand{\p}[1]{{\rm{#1}}}
\newcommand\T{\rule{0pt}{2.6ex}}
\newcommand\B{\rule[-1.2ex]{0pt}{0pt}}
\newcommand{\reffig}[1]{\mbox{Fig.~\ref{#1}}}
\newcommand{\refeq}[1]{\mbox{Eq.~(\ref{#1})}}
\hyphenation{Fesh-bach}
\newcommand{\nuke}[1]{}
\newcommand{\alert}[1]{\textcolor{red}{[\textrm{#1}]}}

\title[rf Feshbach resonances]{Creation and manipulation of Feshbach resonances with radio-frequency radiation}

\author{Thomas M. Hanna, Eite Tiesinga and Paul S. Julienne}
\address{Joint Quantum Institute, 
	NIST and University of Maryland,  
	100 Bureau Drive, Stop 8423,
  	Gaithersburg MD 20899-8423, USA}
\ead{tom.hanna@merton.oxon.org}

\begin{abstract}

We present a simple technique for studying collisions of ultracold atoms in the presence of a magnetic field and radio-frequency radiation (rf).
Resonant control of scattering properties can be achieved by using rf to couple a colliding pair of atoms to a bound state.
We show, using the example of \lisix, that in some ranges of rf frequency and magnetic field this can be done without giving rise to losses.
We also show that halo molecules of large spatial extent require much less rf power than deeply bound states. 
Another way to exert resonant control is with a set of rf-coupled bound states, linked to the colliding pair through the molecular interactions that give rise to magnetically tunable Feshbach resonances.
This was recently demonstrated for \rbs~\cite{kaufman09}.
We examine the underlying atomic and molecular physics which made this possible.
Lastly, we consider the control that may be exerted over atomic collisions by placing atoms in superpositions of Zeeman states, and suggest that it could be useful where small changes in scattering length are required.
We suggest other species for which rf and magnetic field control could together provide a useful tuning mechanism. 
\end{abstract}

\maketitle

\section{Introduction}

Many recent studies of ultracold gases have depended on the manipulation of atomic collisions with external fields.
One way of achieving such control is magnetically tunable Feshbach resonances~\cite{tiesinga93, inouye98, timmermans99}.
Also, lasers have been used to create optical Feshbach resonances by coupling a colliding pair to a bound state of an excited potential~\cite{fedichev96, fatemi00}.
A number of recent works have used optical frequency lasers in combination with magnetically tunable Feshbach resonances to
probe~\cite{courteille98, chin03} and modify~\cite{bauer09} scattering
and bound state properties.
This can provide additional control capabilities, due to advantageous properties of laser beams such as the ready tunability of power and frequency.
Furthermore, some experiments of current interest, such as color superfluidity~\cite{rapp07} and Efimov physics~\cite{dincao09}, could benefit from the ability to independently control scattering lengths between different pairs of a multicomponent gas.

Radiofrequency radiation (rf) is an essential and much used tool in atomic physics, being central to the operation of experiments such as atomic clocks~\cite{audoin}.
In the context of ultracold gases, rf has been used both to dissociate~\cite{regal03} and associate~\cite{thompson05, ospelkaus06, beaufils10} molecules, as well as to drive transitions between bound states~\cite{lang08b}.  
As a probe, rf allows measurements of interaction effects~\cite{gupta03, regal03b}, molecular binding energies~\cite{regal03, thompson05, ospelkaus06}, and the pairing gap of fermionic superfluids~\cite{chin04science}.  
These experiments creating or studying bound states used the rf to create a degeneracy between dressed bound and scattering states, i.e., a Feshbach resonance.
This provides a clear motivation for study of how the same degeneracy can be used to control the collision properties of a pair of atoms.

In Ref.~\cite{kaufman09}, rf control of \rbs\ collisions was demonstrated and studied theoretically. 
Furthermore, rf has been considered as a means for controlling scattering lengths in a variety of scenarios, with a number of different theoretical methods.
Moerdijk \textit{et al.} considered collisions of rf-dressed Na atoms in a magnetic trap using a coupled-channels approach~\cite{moerdijk96}. They showed that favourable conditions for evaporative cooling could be created, and suggested controlling scattering lengths using rf coupling to create a Feshbach resonance.
Zhang \textit{et al.} used a two-channel parametrisation of a magnetically tunable Feshbach resonance, together with atomic rf dressing to tune the resonance location, or multi-frequency rf to independently tune different scattering lengths in a multicomponent gas~\cite{zhang09}.
Tscherbul \textit{et al.} performed a coupled channels analysis of \rbs\ collisions, studying both the creation and manipulation of resonances with rf~\cite{tscherbul10}.
Papoular \textit{et al.} suggested rf control of collisions in gases at zero magnetic field~\cite{papoular09}. 
Lastly, we note the suggestions of Alyabyshev \textit{et al.} to use rf to control atom-molecule interactions~\cite{alyabyshev09}, particularly to suppress inelastic collisions~\cite{alyabyshev09b}.

Studies of ultracold gases have benefited immensely from theoretical
prediction and characterisation of Feshbach resonances~\cite{chin_review}, for which coupled channels calculations are a leading technique~\cite{stoof88, tiesinga93, mies96}. 
These have the drawback of being computationally intense in some cases. 
For this reason, a number of simpler techniques have been developed, including the asymptotic bound state~\cite{wille08, tiecke09}, accumulated phase~\cite{vankempen02}, and three-parameter van der Waals models~\cite{hanna09}.
Our technique~\cite{hanna09} is able to calculate scattering and bound state properties on the basis of only three parameters describing the interactions.
In Ref.~\cite{kaufman09} we extended it to include the effects of rf radiation on collisions of \rbs. 
In this paper we present several new results illustrating the utility of rf for controlling ultracold collisions, and explain our model in greater detail.

Radiofrequency radiation can modify collision properties in three ways, each of which we examine here. 
Firstly, rf can couple a colliding pair and a bound state (bound-free coupling), as used in the above-mentioned experiments on creation and measurement of molecules~\cite{regal03, thompson05, ospelkaus06, beaufils10}. 
This creates a situation similar to an optical Feshbach resonance. 
We consider nonzero magnetic fields, and large halo states created by magnetically tunable Feshbach resonances, giving example results for \lisix. 
We show that this reduces the power required for useful control from that necessary for a deeply bound state. 
In general, rf resonances cause losses by coupling the colliding pair to a large number of energetically lower configurations of the internal atomic states, which we refer to as channels. 
Interestingly, though, we have found some regions of rf frequency and magnetic field in which creating a resonance using a deeply bound state does not lead to losses.
Secondly, rf can couple several bound states together (bound-bound coupling), which can then interact with the colliding pair through the molecular interactions which give rise to magnetically tunable Feshbach resonances.
This was recently demonstrated for \rbs\ in Ref.~\cite{kaufman09}. 
The Franck-Condon factors between two bound states are generally larger than those between bound and scattering states, allowing rf control with comparatively low power.
Our theory was applied to this situation in Ref~\cite{kaufman09}; here, we give a more detailed analysis.
Lastly, dressing an atom with rf creates a superposition of atomic states. 
The collision of two such atoms then involves a superposition of several entrance channels (free-free coupling). 
If the scattering length varies significantly between channels, this alone could be used to vary the scattering length. 

To make this paper self-contained, we start with a description of our three-parameter model in Sec.~\ref{sec:mqdt}.
The main extension necessary to include the effects of rf is the formulation of an appropriate basis for the scattering calculation, which we discuss in Sec.~\ref{sec:rf}.
Our results for bound-free, bound-bound and free-free coupling are presented in Secs.~\ref{sec:li}, \ref{sec:rb} and \ref{sec:liff}, respectively. 
For each, we consider the extent of control that is possible, the rf power required, and the accompanying losses.
We conclude in Sec.~\ref{sec:conclusion}.

%%%%%%%%%%%%%%%%%%%%%%%%%%%%%%%%%%%%
\section{Scattering theory approach}
\label{sec:mqdt}

We consider the collision of two $^{2}$S alkali atoms. 
In the presence of a magnetic field, $\vec{B} = B\hat{z}$, 
the atoms are described by a set of Zeeman states $|\alpha\rangle$ with energies $E_{\alpha}$. 
We label such states $\alpha = a,b,\ldots$,  in order of increasing energy.
These correlate with the zero field states $|f, m_f\rangle$, where $f$ is the total angular momentum of the atom, and $m_f$ its projection along  $\hat{z}$. 
At non-zero field, $m_f$ remains a good quantum number but $f$ does not.
The collision of a pair of atoms is described by an expansion of the interaction Hamiltonian in channels 
$|\alpha_1 + \alpha_2 \rangle |\ell m_\ell \rangle$, 
defined by the state of atoms 1 and 2, the partial wave $\ell$ of their collision, and $m_\ell$, the projection of $\vec{\ell}$ along $\hat{z}$. 
Several coupled channels are typically involved in a collision. 
The threshold energy of a channel is defined as 
$E_{\alpha_1 + \alpha_2} = E_{\alpha_1} + E_{\alpha_2}$, 
i.e. the energy of two atoms in the relevant atomic states at asymptotically large separation $r$, with zero relative kinetic energy.
If the threshold energy of a channel is below the total energy of the colliding atoms, the channel is called open; otherwise, it is referred to as closed.
At zero magnetic field, channels of the same total angular momentum $\vec{T} = \vec{f}_1 + \vec{f}_2 + \vec{\ell}$ are coupled.
At non-zero field, however, channels of the same 
$M_T = m_1 + m_2 + m_\ell$
are coupled.

The valence electrons of the two colliding atoms give rise to two Born-Oppenheimer (BO) interaction potentials, one each of singlet and triplet symmetry.
These are isotropic, independent of $\ell$ and $m_\ell$, and for large $r$ take the form of a van der Waals potential, $-C_6/r^6$.
Here, $C_6$ is the van der Waals coefficient, which is the same for both BO potentials.
Expanding the Hamiltonian in terms of the channels $|\alpha_1 + \alpha_2 \rangle |\ell m_\ell \rangle$, the diagonal matrix elements have the long range form
\begin{equation}
-\frac{C_6}{r^6} + \frac{\hbar^2 \ell(\ell+1)}{2m_\p{r} r^2} + E_{\alpha_1} + E_{\alpha_2} \, ,
\label{eq:Vlongrange}
\end{equation}
where $m_\p{r}$ is the reduced mass. 
Off-diagonal elements of the BO potentials decay exponentially as $r$ increases. 
These can lead to inelastic spin relaxation (ISR), the loss of atoms by decay into an energetically lower channel.
Such losses are also referred to as inelastic spin-exchange collisions.
Weaker effects such as relativistic spin-dependent interactions~\cite{chin_review} are not included in the present calculations.
The potential takes the form of \refeq{eq:Vlongrange} for $r > r^{*}$, where $r^*$ is a distance at which the splittings between diagonal elements have the same order of magnitude as the off-diagonal elements.
This distance is usually of order $20\,a_0$, where $a_0 = 0.05292$\,nm is the Bohr radius.

For $r < r^*$, the depths of and splittings between the singlet and triplet BO potentials are much larger than  the relative kinetic energy of the colliding atoms and the atomic hyperfine splittings.
This energy-scale separation leads to the concept of a quantum defect, in which a simple parametrisation is used to account for the short-range physics, and is matched with solutions of the long-range potential. 
These long-range solutions are much easier to obtain. 
Such an approach is valid because the short range region is not probed in detail by the colliding pair.
We note that quantum defect theory has been used in a wide variety of contexts including atomic scattering~\cite{mies84, gao05}, electron-ion collisions and Rydberg states~\cite{greene82, seaton83}, and nucleon scattering~\cite{bethe35}.

Our approach is based on a series of papers by Gao  (see Refs.~\cite{gao96, gao98, gao05, gao08} and references therein), in which a number of powerful tools for pure $C_n/r^n$ potentials were presented.
For each channel we define a reference potential, taken to be \refeq{eq:Vlongrange} extended over all $r$. 
From this we calculate reference functions  $f$ and $g$, which are two linearly independent solutions of the Schr\"odinger equation.
We use these to form vectors $\vec{f}$ and $\vec{g}$ spanning all channels.
In the original problem, mixing between channels occurred due to spin exchange. 
Here, this mixing is incorporated in the $r = 0$ boundary condition.
As done in the work of Gao~\cite{gao05}, by choosing $f$ and $g$ appropriately we can express this boundary condition as a short range $K$ matrix, $\mb{K}^{(s)}$, that is independent of collision energy $E$ and partial wave. 
Here, we use bold font to indicate a matrix.
In fact, the multichannel wavefunction with our approximate Hamiltonian can be written as
\begin{equation}
 \vec{\psi}(r) = \vec{f}(r) - \mb{K}^{(s)} \vec{g}(r) \, ,
 \label{eq:psiofk}
\end{equation}
for all $r$. 

A convenient molecular basis for calculating $\mb{K}^{(s)}$ is that  described by
$|(s_1 s_2)S(i_1 i_2)I ; F \ell ; T M_T\rangle $,
where
$s_{1,2}$ and $i_{1,2}$ are the electronic and nuclear spin angular momentum of atoms 1 and 2, respectively.
The two electron spins are coupled together, as are the two nuclear spins, to give the total electron spin $\vec{S} = \vec{s}_1 + \vec{s}_2$ and total nuclear spin $\vec{I} = \vec{i}_1 + \vec{i}_2$.
These are then coupled to give $\vec{F} = \vec{S} + \vec{I}$, which is coupled to the partial wave $\vec{\ell}$ to give the total angular momentum, $\vec{T}$.
In this basis, $\mb{K}^{(s)}$ is diagonal, with diagonal entries depending only on whether the corresponding channel is of singlet ($S=0$) or triplet ($S=1)$ symmetry.
Their values $K_\p{s,t}$ are given by the relation~\cite{gao05}
\begin{equation}
  a_{\p{s,t}} / \bar{a} = \sqrt{2} 
  \frac{K_{\p{s,t}} + \tan(\pi/8)}{K_{\p{s,t}} - \tan(\pi/8)} \, .
  \label{eq:ast}
\end{equation}
Here, $a_\p{s,t}$ are the scattering lengths of the singlet and triplet potentials, respectively, 
$\bar{a} = 2^{-1/2}[\Gamma(3/4)/\Gamma(5/4)]\, l_\p{vdW}$ is the mean scattering length, 
$l_\p{vdW} = (2 m_\p{r} C_6/\hbar^2)^{1/4}/2$ is the van der Waals length, and $\Gamma(z)$ is the gamma function. 
We use a frame transformation~\cite{fano70, rau71} to convert the $K$ matrix to the basis $\ket{\alpha_1 + \alpha_2}\ket{\ell m_\ell}$.
As these are not associated with singlet or triplet symmetry, we then have off-diagonal terms in $\mb{K}^{(s)}$.

Determination of scattering properties is based on the calculation of the reference functions $f$ and $g$.
The Schr\"odinger equation with a pure van der Waals potential can be solved analytically~\cite{gao05}.
At short range, $f$ and $g$ have the limiting form
\begin{eqnarray}
 f(s)  = & \frac{1}{2\sqrt{\pi}} s^{3/2} \cos (\sqrt{2} s^{-2} - \pi/4) \, , \\
 g(s) = & -\frac{1}{2\sqrt{\pi}} s^{3/2} \sin (\sqrt{2} s^{-2} - \pi/4) \, .
\end{eqnarray}
Here, we have defined $s = r / l_\p{vdW}$. These functions can be recognised as the zero-energy Wigner-Kramer-Brillouin solutions to the van der Waals potential. For open channels, they are linked to the long range forms, 
\begin{eqnarray}
f(r) &= \frac{1}{\sqrt{\pi k l_\p{vdW}}}
\left[ Z_{fs}(E) \sin\left(kr - \frac{\ell \pi}{2}\right) 
- Z_{fc}(E) \cos \left(kr - \frac{\ell\pi}{2}\right) \right] \, , 
\label{eq:fl}
\\
g(r) &= \frac{1}{\sqrt{\pi k l_\p{vdW}}}
\left[ Z_{gs}(E)\sin \left(kr - \frac{\ell \pi}{2}\right)
 - Z_{gc}(E) \cos \left(kr - \frac{\ell\pi}{2}\right) \right] \, .
 \label{eq:gl}
\end{eqnarray}
Here, $k$ is the wavenumber of the relative motion.
We note that the $Z$ coefficients are also $\ell$ dependent.
For closed channels, the reference functions become analogous superpositions of exponentially growing and decaying functions.

The reference functions allow us to find the physical $\mb{K}(E)$ matrix for the open channels, from which scattering properties can be extracted.
This takes the form~\cite{gao05}
\begin{equation}
  \mb{K}(E) = - [\mb{Z}_{fc}(E) - \mb{Z}_{gc}(E) \mb{K}_\p{eff} ]  [\mb{Z}_{fs}(E) - \mb{Z}_{gs}(E) \mb{K}_\p{eff} ]^{-1} \, ,
  \label{eq:Keff}
\end{equation}
where
\begin{equation}
 \mb{K}_\p{eff} = \mb{K}_{oo}^{(s)}+ \mb{K}_{oc}^{(s)} [\boldsymbol{\chi}(E) - \mb{K}_{cc}^{(s)}]^{-1} \mb{K}_{co}^{(s)}
\, .
\label{eq:Keffz}
\end{equation}
Here, the `oo' and `cc' refer to the open and closed channel blocks of $\mb{K}^{(s)}$, while `oc' and `co' indicate the open-closed blocks. 
The $Z$ matrices are diagonal and have entries equal to the $Z$ coefficients of Eqs.~(\ref{eq:fl}) and (\ref{eq:gl}) evaluated for the corresponding channels.
The diagonal matrix of the bound state phase in the closed channels, $\boldsymbol{\chi}(E)$, is analogously found from the long range behaviour of the reference functions in the closed channels. 
It allows the calculation of bound state energies from the determinental equation~\cite{gao05}
\begin{equation}
 \p{det}( \boldsymbol{\chi}(E) - \mb{K}_{cc}^{(s)} ) = 0 \, .
 \label{eq:bnd_det}
\end{equation}

We are able to extract all desired scattering properties from $\mb{K}(E)$. We first calculate the $S$ matrix, $\mb{S}(E) = [1 + i\mb{K}(E)] [1 - i\mb{K}(E)]^{-1}$. 
In the presence of just one open channel and the limit of $k \rightarrow 0$, the scattering length can then be found from the relation
\begin{equation}
S(E) = \exp(-2ika) \, .
\label{eq:sofe}
\end{equation}
We note that $\mb{S}(E)$, which is only defined for open channels, is a scalar for this case.
Equation~(\ref{eq:sofe}) can still be used when there are several open channels. 
However, the diagonal $S$ matrix element of the entrance channel will have less than unit modulus. 
We can then reinterpret the right side of \refeq{eq:sofe} as 
$\exp(-2ik\tilde{a})$, 
where
$\tilde{a} = a - ib$
is the complex scattering length, and $a$ and $b$ are real.
For any collision energy, the two-body decay rate coefficient is
\begin{equation}
K_2 = \frac{\pi\hbar}{m_\p{r} k}\sum_{i \neq e} |S_{ei}(E)|^2 \, ,
\end{equation}
where the index $i$ ranges over all open channels other than the entrance channel $e$. 
For the multichannel case, an alternative way of finding $b$ is to extract it from $K_2$ in the limit $k \rightarrow 0$, using the formula
\begin{equation}
b = \frac{m_\p{r}}{2 h} K_2 \, .
\label{eq:K2b}
\end{equation}

Our model can be optimised to measured data by using the singlet and triplet scattering lengths as fit parameters~\cite{hanna09}. 
This can be used for prediction of further resonances, or just to offset the limitations of our simplified approach.
These limitations were discussed in Ref.~\cite{hanna09}. 
Briefly, resonances arising as a result of deeply bound states can be predicted only approximately, as non-van der Waals parts of the potential are significant. 
For bound states accurately reproduced by the $-C_6/r^6$ potentials, however, our approach is accurate. 
The examples considered in the present work fall within this category.
With the approximations of using a van der Waals potential and assuming energy independence at short range, our approach can predict all observable scattering properties from known atomic parameters and three properties of the interactions: $a_\p{s}$, $a_\p{t}$ and $C_6$.

%%%%%%%%%%%%%%%%%%%%%%%%%%%%%%%%%%%%%%%%%%%%
\subsection{Decaying resonances}
\label{sec:decay}

Resonances invariably create some losses in ultracold collisions. Inelastic spin relaxation (ISR), spin-spin dipole coupling between partial waves, and three-body recombination are all enhanced near a Feshbach resonance.
As shown in \refeq{eq:K2b}, two-body decay into energetically lower exit channels leads to an imaginary part of the scattering length.
The losses in our calculations represent ISR.
The energy gap between channels is typically large enough that atoms undergoing ISR are lost from the system.
Our calculations do not include three-body recombination or decay into other partial waves.

Theoretical work on decaying resonances has focused on the optical case, where a laser couples a colliding pair to an excited bound state which can spontaneously decay.
This theory can be readily adapted to a magnetically tunable decaying resonance, or one in which both a magnetic field and rf are used.
Following Bohn and Julienne~\cite{bohn99}, we can write the complex scattering length in the limit $k \rightarrow 0$ as
\begin{eqnarray}
a(B) &= &a_\p{bg}  \left(1 - 
\frac{\Delta(B - B_0)}
	{(B - B_0)^2 + (\gamma_\p{B}/2)^2} \right) \, , 
	\label{eq:aofb}
	\\
b(B) &= &2 a_\p{res} 
\frac{(\gamma_\p{B}/2)^2}
{(B - B_0)^2 + (\gamma_{\rm{B}}/2)^2} \, .
\label{eq:bofb}
\end{eqnarray}
Here, $a_\p{bg}$ is the background scattering length, representing the scattering length of the entrance channel in the absence of a resonance. 
We have expressed the decay rate of the bound state, $\gamma$, in magnetic field units, $\gamma_\p{B} = \hbar \gamma/\mu_\p{res}$, where $\mu_\p{res}$ is the difference in magnetic moment between the entrance channel and the bound state causing the resonance.
The resonance length $a_\p{res}$ is defined by $a_\p{res}\gamma_\p{B} = a_\p{bg} \Delta$.
The width and magnetic field location of the resonance are given by $\Delta$ and $B_0$, respectively. 
In the limit of $\gamma \rightarrow 0$ the above formulas reduce to the standard relation describing the scattering length around a non-decaying resonance:
\begin{equation}
a(B) = a_\p{bg}\left( 1 - \frac{\Delta}{B - B_0} \right) \, .
\end{equation}

%%%%%%%%%%%%%%%%%%%%%%%%%%%%%%%%%%%
\section{rf-dressed basis}
\label{sec:rf}

\begin{figure}[tb]
	\centering
	\includegraphics[width=0.6\columnwidth, clip]{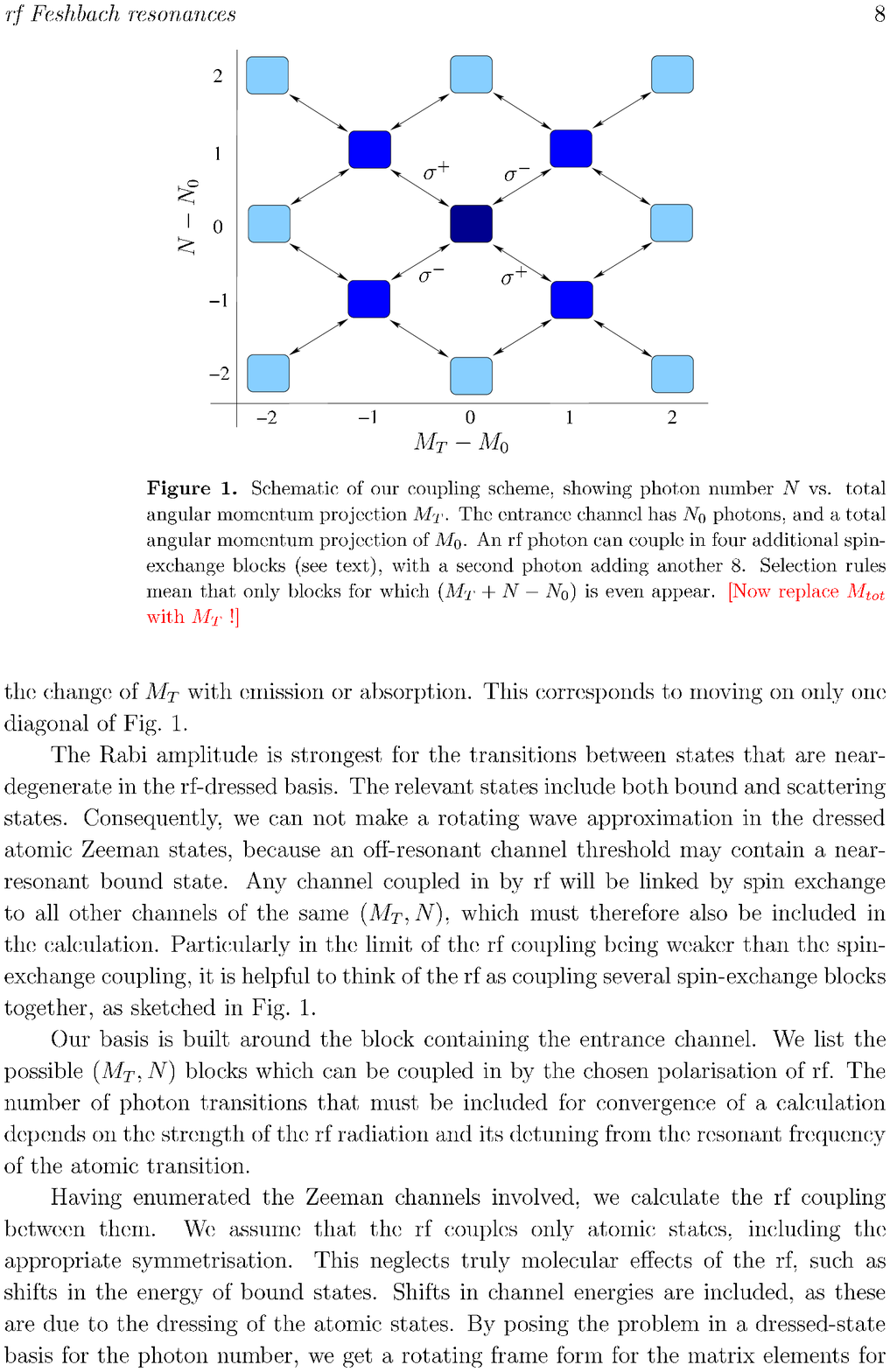}
	\caption{Schematic of our coupling scheme, showing photon number $N$ vs. total angular momentum projection $M_T$. 
	Boxes represent groups of rf-dressed Zeeman channels with the same $(M_T, N)$, which we call spin-exchange blocks.
	The entrance channel is in the central $(M_0, N_0)$ block.
	Arrows indicate rf coupling. 
	A $\sigma_x$ rf photon can couple in four additional spin-exchange blocks, with a second photon adding another 8. 
	Circular polarised light, by contrast, constrains the system to one diagonal, as indicated by the $\sigma^{\pm}$ labels.}
	\label{fig:coupling}
\end{figure}

Radio-frequency radiation drives transitions between atomic Zeeman states.
In the two-body picture, rf couples Zeeman channels together.
We therefore use a basis of rf-dressed channels,  
$\ket{\alpha_1 + \alpha_2, N}$, 
where $N$ is the number of photons in the rf field. 
In the following sections we consider only $s$-wave collisions, and so omit the partial wave  labels.
In \reffig{fig:coupling} we show a schematic of our coupling scheme.
All channels with the same $M_T$ are coupled together by spin exchange, the molecular interactions which give rise to magnetically tunable Feshbach resonances and ISR in the absence of rf.
We refer to a set of channels with the same $(M_T, N)$ as a spin-exchange block, indicated by a box in \reffig{fig:coupling}.
Arrows indicate rf coupling.
It is usually possible to identify the entrance channel in which the gas and rf field are initially prepared, and for which we set $M_T = M_0$ and $N = N_0$.

The polarisation of the rf field significantly alters the collisional and loss properties by determining which blocks couple.  
Light that has $\sigma_x$ polarisation, i.e. linear and perpendicular to the magnetic field quantisation axis, drives $(\Delta M_T = \pm 1, \Delta N = \pm 1)$ transitions, as sketched in \reffig{fig:coupling}. 
Here, the two `$\pm$' signs are independent - that is, both an increase and decrease in $M_T$ can be achieved with either an absorption or an emission of an rf photon. 
As a consequence, $\sigma_x$ rf will always couple energetically lower exit channels, and so create losses. 
This can be seen from \reffig{fig:coupling}: if a channel \ket{\alpha+\beta, N} is part of the basis, it is coupled to \ket{\alpha+\beta, N-2}.
When a transition from \ket{\alpha+\beta, N} to \ket{\alpha+\beta, N-2} takes place, $2\hbar\omega$ of energy is transferred from the rf field to the kinetic energy of the atom pair.
This is typically enough energy for the atoms to be lost from the trap.
Circularly polarised light, however, will constrain the sign of $\Delta M_T$ with emission or absorption. 
This corresponds to moving on only one diagonal of \reffig{fig:coupling}, which makes it possible to create a situation in which a pair of atoms in the entrance channel have no allowed exit channels.
For $\sigma^{\pm}$ and $\sigma_x$ light, the change in $M_T$ with absorption or emission means that only spin-exchange blocks for which $(M_T - M_0) + (N - N_0)$ is even appear.

We can not make a rotating wave approximation in the rf-dressed channel energies, because channels with off-resonant energies can contain a near-resonant bound state. 
To make sure no such bound states are omitted, we begin with the block containing the entrance channel and add all blocks that can be coupled by the chosen rf polarisation, up to the required number of rf transitions. 
The number of rf transitions that must be included for convergence of a calculation depends on the two-body spectrum, as well as the strength of the rf radiation and its detuning from the atomic transition frequencies.
The rf coupling between Zeeman channels is taken to be that between the constituent atomic states, including the appropriate two-body symmetrisation. 
The rf adds a term to the Hamiltonian of the form 
$H_\p{rf} = -(\vec{\mu}_1 + \vec{\mu}_2)\cdot \vec{B}_\p{rf}$, where $\vec{\mu}_{1,2}$ are the magnetic moments of atoms 1 and 2, and $\vec{B}_\p{rf}$ is the rf magnetic field operator.
The two-body matrix element can be easily formed from the matrix elements between rf-dressed atomic Zeeman states $\ket{\alpha, N}$.
For $\sigma_x$ radiation, the atomic matrix element is
\begin{eqnarray}
& \langle \alpha, N | -\vec{\mu}\cdot \vec{B}_\p{rf} | \alpha',N' \rangle
& = -\delta_{N, N'\pm 1} \frac{B_\p{rf}}{2} \left(\mu_e\langle \alpha | S_x | \alpha'\rangle 
+ \mu_n \langle \alpha | I_x | \alpha' \rangle \right)
\, .
\label{eq:rfelement}
\end{eqnarray}
Here, $B_\p{rf}$ is the amplitude of the oscillating field, the $N$ dependence of which will be neglected since $N$ is assumed to be large, $S_x = (S_{+1} - S_{-1})/\sqrt{2}$, and $\mu_{e, n}$ are the magnetic moments of the electronic state and the nucleus, respectively. 
For $\sigma^{\pm}$ polarised light, $S_{x}$ and $I_x$ must be replaced with $S_{\pm 1}$ and $I_{\pm 1} $. 
The remaining matrix elements are calculated by decomposing the Zeeman states into the components of the electron spin $s$ and nuclear spin $i$, and realising that
\begin{equation}
\langle s' m_s' | S_q | s m_s \rangle = \delta_{s' s} \delta_{m_s', (m_s+q)} (-1)^{q} \sqrt{s(s+1)} C(s 1 s; m_s + q, -q, m_s) \, .
\end{equation}
Here, $C$ is a Clebsch-Gordan coefficient, and $q = -1, 0,$ or 1~\cite{rose}. 
The matrix element for $I_q$ can be calculated in the same manner. 
In this paper, we find it convenient to express the strength of the rf coupling in terms of the Rabi frequency between the two energetically lowest Zeeman states, with the pertinent polarisation:
\begin{equation}
h \Omega = 
2 \,|\langle a, N_0 | -\vec{\mu}\cdot \vec{B}_\p{rf} | b,N_0-1 \rangle | \, .
\end{equation}
The Rabi frequencies for all other atomic transitions then follow from angular momentum algebra. 
This definition is weakly dependent on the bias field $B$, however its variation over the field ranges shown in our examples is negligible. 

\begin{figure}[tb]
	\centering \includegraphics[width=0.6\columnwidth,
		clip]{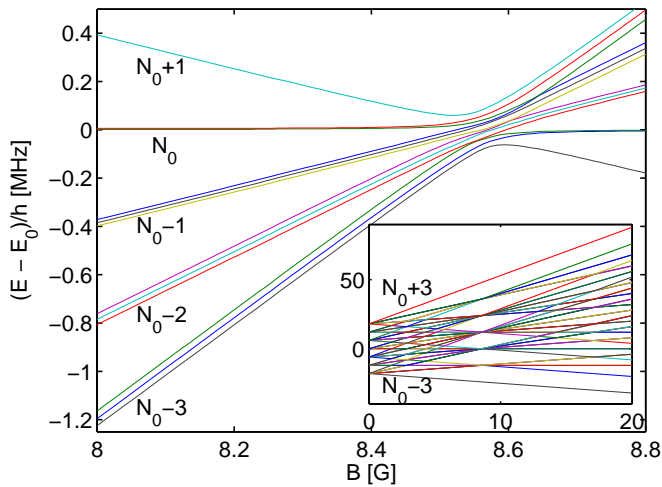}
	\caption{Radio-frequency dressed channel energies of 
	two \rbs\ atoms	as a function of magnetic field. 	
	Energies are given relative to $E_0$, the energy of the $|a + e, N_0 \rangle$ entrance channel in the absence of coupling to other channels. 	
	Avoided crossings between many of these states occur near 8.6\,G.
	The inset shows the same quantities over a wider range. 
	Groups of channels are labelled with the number of photons associated with them.
	Here, the rf oscillation frequency is 6\,MHz, the Rabi frequency is 38\,kHz, and $1\,\rm{G}= 10^{-4}\,\rm{T}$.
	}
	\label{fig:rfen_2b}
\end{figure}
In extending the method of Sec.~\ref{sec:mqdt} to include rf, we add to the approximations that the rf coupling is negligible at short range. 
That is, the matrix $\mb{K}^{(s)}$ is still independent of energy, and is also diagonal in $N$.
In order to use the established tools of scattering theory, a basis must be chosen that makes the Hamiltonian diagonal at asymptotically large $r$.
For the present case, a basis of two-body Zeeman channels will have off-diagonal rf coupling matrix elements, as well as Zeeman energies on the diagonal. 
We therefore explicitly diagonalise the Hamiltonian for the case $r \rightarrow \infty$, which 
provides a basis of rf coupled states. 
The eigenvectors can then be used to express the scattering matrix in this basis.
An example of dressed \rbs\ channel energies for magnetic fields around 8.6\,G and rf frequency of 6\,MHz are shown in \reffig{fig:rfen_2b}.
For this example, which we discuss at length in Sec.~\ref{sec:rb}, the Zeeman effect is mainly linear, leading to several nearby avoided crossings between channels with different photon numbers.
In \reffig{fig:rfen_2b} we have only shown the channels corresponding to the $B = 0$ limit $(f_1 = 1) + (f_2 = 2)$. 
However, our calculations include all Zeeman channels of the relevant $(M_T, N)$ blocks.

%%%%%%%%%%%%%%%%%%%%%%%%%%%%%%%%%%%%%%%
\section{Results}
\label{sec:examples}

The technique we have developed is general and can be used to examine any pair of $^{2}$S alkali atoms. In this section we present three significantly different examples.
In Sec.~\ref{sec:li} we consider directly creating resonances by rf coupling a colliding pair to a bound state, focussing on the example of \lisix. 
In Sec~\ref{sec:rb} we discuss \rbs, in which several bound states are coupled together by rf. 
This bound state manifold is then coupled to the colliding pair by molecular interactions.
Lastly, we consider the control that can be obtained with rf dressing of atomic Zeeman states of \lisix\ in Sec.~\ref{sec:liff}. 

%%%%%%%%%%%%%%%%%%%%%%%%%%%%%%%%%%%%%%%
\subsection{{\lisix}: bound-free coupling}
\label{sec:li}

The broad $s$-wave resonances in the $a + b$, $a+c$ and $b+c$ channels  of \lisix\ have been used for studying effects such as the BEC-BCS crossover~\cite{zwierlein04, kinast04, bartenstein04} and Efimov physics~\cite{ottenstein08, huckans09, williams09}. 
Three-component \lisix\ gases (with atoms in Zeeman states $a$, $b$, and $c$) near these broad, overlapping resonances have also been considered as candidates for observing color superfluidity~\cite{rapp07}.
This and other applications could benefit from a second degree of control to allow tuning of the interactions between different component pairs.
In this subsection we consider rf resonances in which the colliding pair of atoms is coupled to a bound state. 
The strongest bound-free Franck-Condon overlaps are provided by halo states.
Substantial control of scattering properties is then possible with less power than is required for a deeply bound state.
However, for some bound states there exist ranges of rf frequency and magnetic field for which the entrance channel is the energetically lowest.
This requires the use of circularly polarised rf, but allows control without the creation of losses. 

\begin{figure}[tb]
	\centering
	\includegraphics[width=0.49\columnwidth, clip]{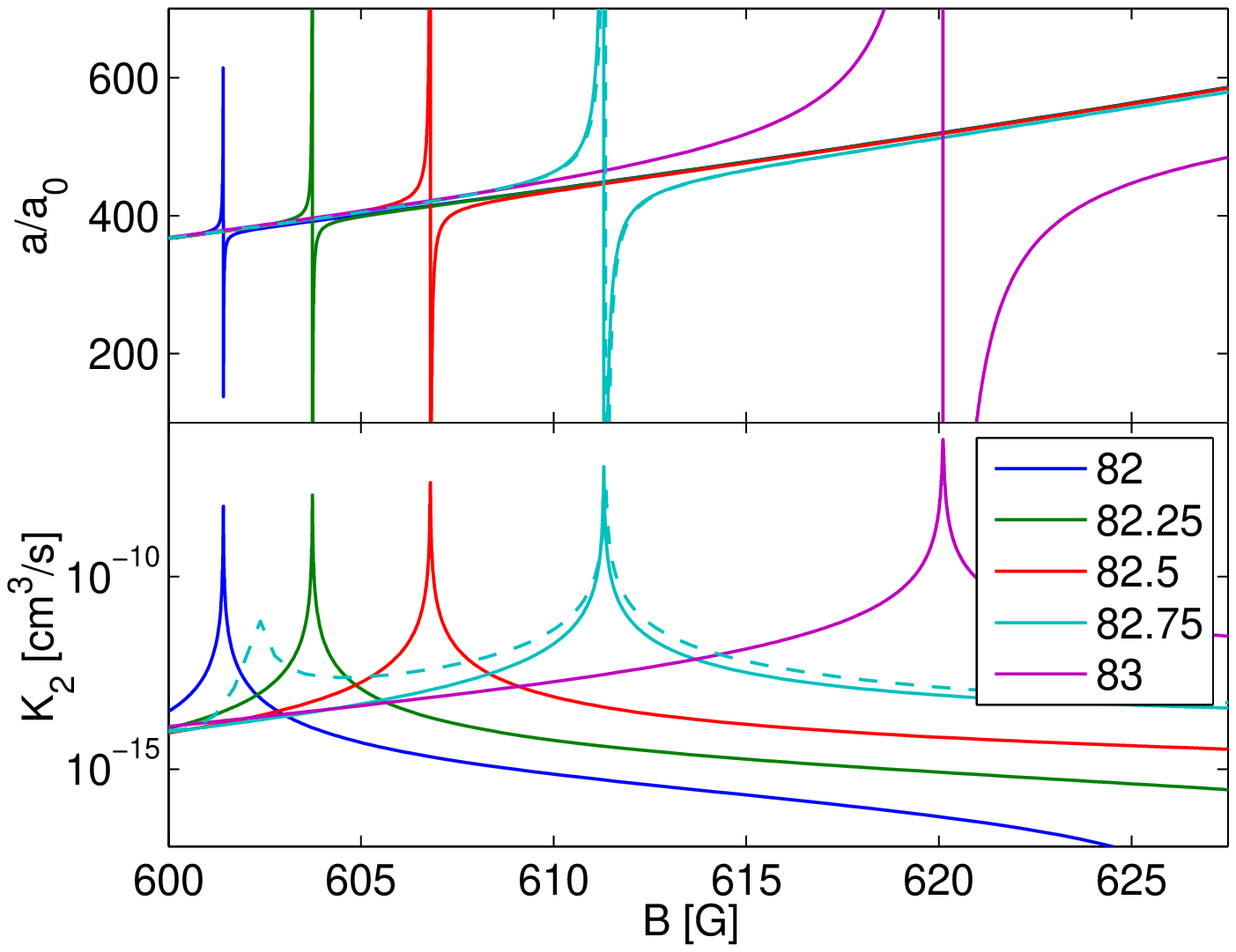}
	\includegraphics[width=0.49\columnwidth, clip]{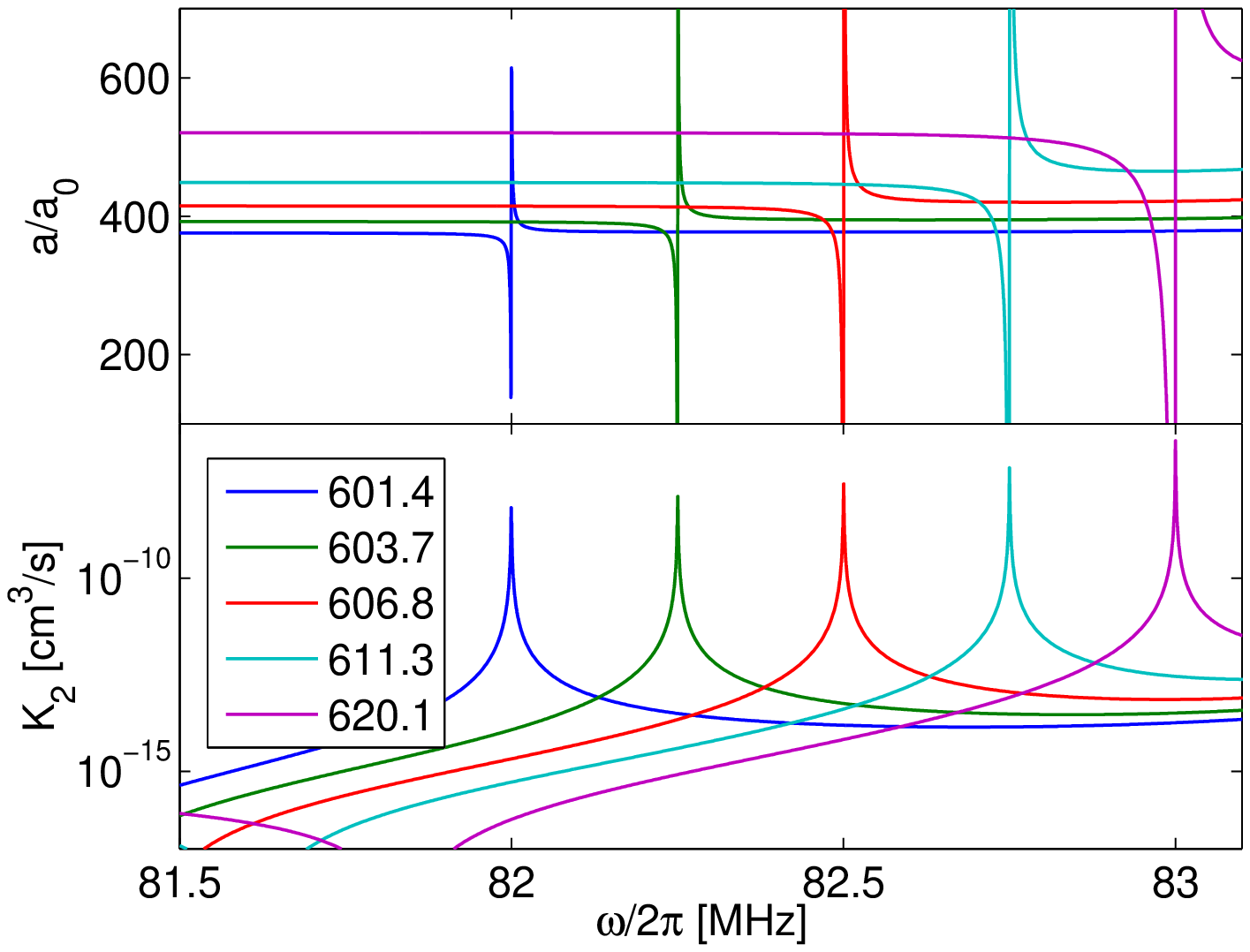}
	\caption{Real part of the scattering length (top) and two-body loss rate coefficient (bottom) for rf induced resonances in \lisix. 
	On the left plot, these are shown as a function of magnetic field, with fixed rf frequencies indicated in MHz. 
	On the right plot, these are shown as a function of rf frequency, with fixed magnetic fields given in G. 
	Lines of the same colour correspond to the same resonance, with the five values of rf frequency / magnetic field in the two plots chosen to pass through the resonance peak.
	The Rabi frequency of the $\sigma^{-}$ rf is 100\,kHz, and the entrance channel is \ket{a+b, N_0}. 
	The dashed line shows the effect of using $\sigma_x$ rf of the same modulation amplitude.
	The resonance positions coincide with the energy of the bound state causing the resonance in the $\ket{a+c, N_0 - 1}$ channel at 690\,G.
	The width of the resonance increases as the bound state moves closer to threshold, enhancing the Franck-Condon overlap of bound and scattering states.}
	\label{fig:li6_halo}
\end{figure}
We show an example of an rf-induced resonance with a halo state in \reffig{fig:li6_halo}.
This resonance is created in the scattering of \lisix\ atoms in the \ket{a+b, N_0} channel, with rf coupling the colliding pair to the bound state that causes a resonance in the energetically closed \ket{a+c, N_0 - 1} channel at 690\,G.
Here, we have used a Rabi frequency of 100\,kHz, corresponding to an oscillation amplitude of 0.5\,G.
The frequency of the $\sigma^{-}$ rf is allowed to vary from 82\,MHz to 83\,MHz, which resonantly couples the bound state at magnetic fields in the range 600\,G -- 620\,G.
The two panels of \reffig{fig:li6_halo} show that the resonance can be tuned with rf frequency or magnetic field.
A larger rf frequency resonantly couples the bound state at a higher magnetic field, for which its energy is closer to the \ket{a+c, N_0-1} threshold. 
The resonance created is then wider as a function of both $B$ and $\omega$, as well as providing a larger variation in scattering length and a higher peak loss rate.
These effects are consequences of the bound state having a more halo-like character, and a larger bound-free overlap with scattering states of all open channels.
These rf-coupled exit channels are the only available means of two-body decay, since in the absence of rf $a+b$ is the lowest channel with $M_T = 0$.

As can be seen from Eqs.~(\ref{eq:aofb}) and (\ref{eq:bofb}), the extrema of the scattering length in the vicinity of such a decaying resonance are $a_\p{bg} \pm a_\p{res}$.
The resonance length $a_\p{res}$ thus provides a convenient indication of how significantly the scattering length can be controlled with a given resonance.
For $\omega/2\pi = 82$\,MHz and $B_0 = 601.4$\,G we have $a_\p{res} = 280\,a_0$, whereas for $\omega/2\pi = 83$\,MHz and $B_0 = 620.1$\,G we have $a_\p{res} = 13000\,a_0$.
The use of circularly polarised rf can reduce the number of available exit channels, and therefore reduce total losses.
The dashed lines in \reffig{fig:li6_halo} illustrate this for the present case, giving the scattering properties for $\sigma_x$ rf of the same modulation amplitude as the solid $\sigma^{-}$ lines.
However, the loss rates could still be too high for experimentally relevant densities.

\begin{figure}[tb]
	\centering
		\includegraphics[width=0.8\columnwidth, clip]{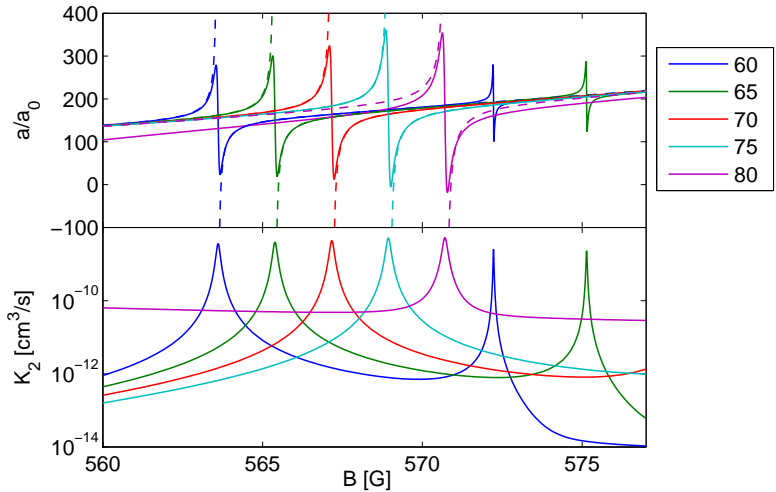}
	\caption{Scattering length (top panel) and two-body loss rate coefficient (bottom panel) as a function of magnetic field, with \ket{a+b, N_0} entrance channel, a Rabi frequency of 1\,MHz and rf frequencies indicated in MHz. 
	The rf resonances are created by coupling to a deeply bound state. 
	While $\sigma_x$ rf (solid lines) produces strong losses, as shown in the bottom panel, for $\sigma^{+}$ (dot-dashed lines) the entrance channel is the energetically lowest state. 
	Consequently, no losses are created by the rf. 
	The small features at the right of the plot are due to a $\sigma^{-}$ transition, and are therefore only present for the $\sigma_x$ rf.
	}
	\label{fig:li6_deep}
\end{figure}

For some ranges of magnetic field strength and rf frequency 
circularly polarised rf allows the coupling in of a bound state without creating losses. 
One such example is shown in \reffig{fig:li6_deep}.
A deeply bound $M_T = 1$ state supported by higher channels is degenerate with the $a+b$ collision threshold close to 543\,G.
In the vicinity of this magnetic field, it may be coupled to the \ket{a+b, N_0} entrance channel using $\sigma^{+}$ radiation, without opening any exit channels.
This is illustrated by the dot-dashed lines in \reffig{fig:li6_deep}. 
The divergence in scattering length characteristic of non-decaying resonances is observed. 
In contrast to this, $\sigma_x$ rf of the same frequency and power produces a decaying resonance, shown by the solid lines in \reffig{fig:li6_deep}.
A maximum in these losses is observed, with a background loss rate that increases as the rf frequency is brought closer to the atomic transition frequency.
Only a slight difference in the width of the resonance is seen as the rf frequency is varied within the range shown, as the properties of the bound state are only weakly varying.
This calculation uses $\Omega/2\pi = 1$\,MHz, corresponding to a modulation amplitude of 5\,G.
Much lower Rabi frequencies could be sufficient, provided that the control of the bias magnetic field is good enough.
The frequency of the rf may be used to decide on the location of the resonance for magnetic field tunability, and the intensity used to choose the strength, analogous to an optical Feshbach resonance. 

\begin{figure}[tb]
	\centering
		\includegraphics[width=0.6\columnwidth, clip]{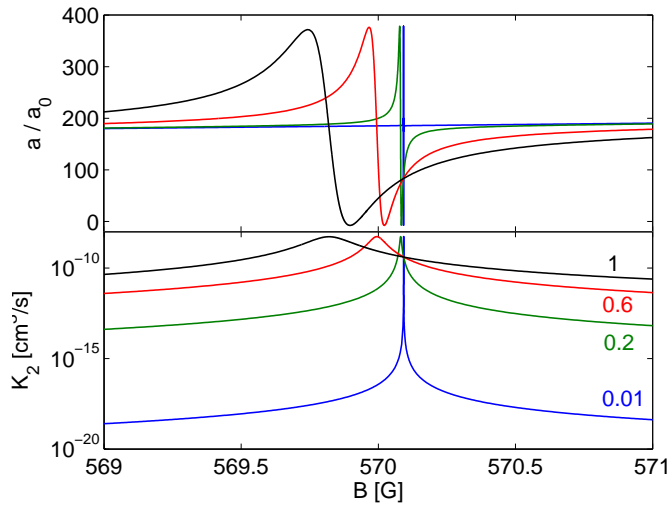}
	\caption{Real part of the scattering length (upper panel) and two-body loss rate coefficient (lower panel) as a function of magnetic field.
	Here, we use the \ket{a+b, N_0} entrance channel and $\sigma^x$ polarised rf with an oscillation frequency of 77.5\,MHz to create a resonance near 570\,G.
	The Rabi frequencies used are given in MHz.
	The width of the loss feature increases strongly with Rabi frequency, as does the background loss created by rf coupling to lower exit channels. 
	However, the maximum available change in the scattering length remains constant. 
	}
	\label{fig:dif_rabi_570}
\end{figure}

As we discussed in Sec.~\ref{sec:rf}, creating a resonance with $\sigma_x$ rf always gives rise to losses.
However, the ability to control the location and width of the loss feature could make such resonances useful as a knife for evaporative cooling.
This was suggested for magnetically tunable resonances by Mathey \textit{et al.}~\cite{mathey09}. 
They found that the width of a resonance limits the temperature to which it can be used to cool a gas.
While the width of a magnetically tunable resonance is set by molecular properties, the width of an rf resonance can be controlled by changing the rf power.
Figure~\ref{fig:dif_rabi_570} shows the scattering length and loss rate coefficient as functions of magnetic field, for Rabi frequencies up to 1\,MHz.
Here, we have used an oscillation frequency of 77.5\,MHz and $\sigma_x$ polarisation.
The Rabi frequency does not alter the nature of the resonantly coupled bound state, and therefore does not change the maximum variation in scattering length.
For each of the resonances in \reffig{fig:dif_rabi_570} we calculate a resonance length of approximately 190\,$a_0$, close to the background scattering length of $184\,a_0$.
However, increasing the Rabi frequency increases the width of the resonance feature. 
For cooling, the width of the resonance could be reduced as the temperature is lowered. 
The lowest achievable temperature would then be set by technical considerations such as magnetic field control.	

The loss properties of an optical or magnetically tunable Feshbach resonance are set by the decay rate of the bare bound state.
This is typically almost constant in the vicinity of the resonance.
In the optical case, this allows the stimulated coupling to be strengthened by increasing the laser power, increasing the resonance length.
In our calculations, the losses are also created by stimulated coupling - i.e., $\gamma_\p{B}$ in  Eqs.~(\ref{eq:aofb}) and (\ref{eq:bofb}) now has the same dependence on Rabi frequency as $\Delta$.
This is why $a_\p{res}$ for the resonances in \reffig{fig:dif_rabi_570} is independent of Rabi frequency.
However, as shown in \reffig{fig:li6_halo}, a significant change in the bound state wavefunction can lead to a strong change in $a_\p{res}$.

%%%%%%%%%%%%%%%%%%%%%%%%%%%%%%%%%%%%%%%%%%%%%%%

\subsection{{\rbs}: bound-bound coupling}
\label{sec:rb}

\rbs\ has been used in a wide range of studies in ultracold gases. 
A large number of resonances have been observed in this system~\cite{marte02}, with many studies utilising the comparatively wide ($\Delta = 0.21$\,G) resonance in the $a+a$ channel at $B_0 = 1007$\,G~\cite{duerr04pra}.
There are a number of resonances grouped close to each other around 9\,G and 18\,G, in channels corresponding to the zero field $(f_1 = 1) + (f_2 = 2)$ limit~\cite{kaufman09, widera04, erhard04, vankempen02}.
These resonances were used in the demonstration and theoretical analysis of rf-dressed Feshbach resonances in Ref.~\cite{kaufman09}. 
We also note the calculations presented in Ref.~\cite{tscherbul10}.
In Ref.~\cite{kaufman09}, the primary effect of the rf was to couple the bound states causing each of the nearby resonances. 
Bound-bound coupling has a substantially larger Franck-Condon overlap than bound-free coupling, and so significant control of scattering properties can be achieved with correspondingly less rf power.
For the calculations shown in this section, the Rabi frequency on the $a \leftrightarrow b$ transition is 38\,kHz, corresponding to $B_{\p{rf}} = 0.08$\,G.

%%%%%%%%%%
\begin{figure}[tb]
	\centering
	 \begin{minipage}[h]{0.49\columnwidth}
\includegraphics[width=0.99\columnwidth, clip]{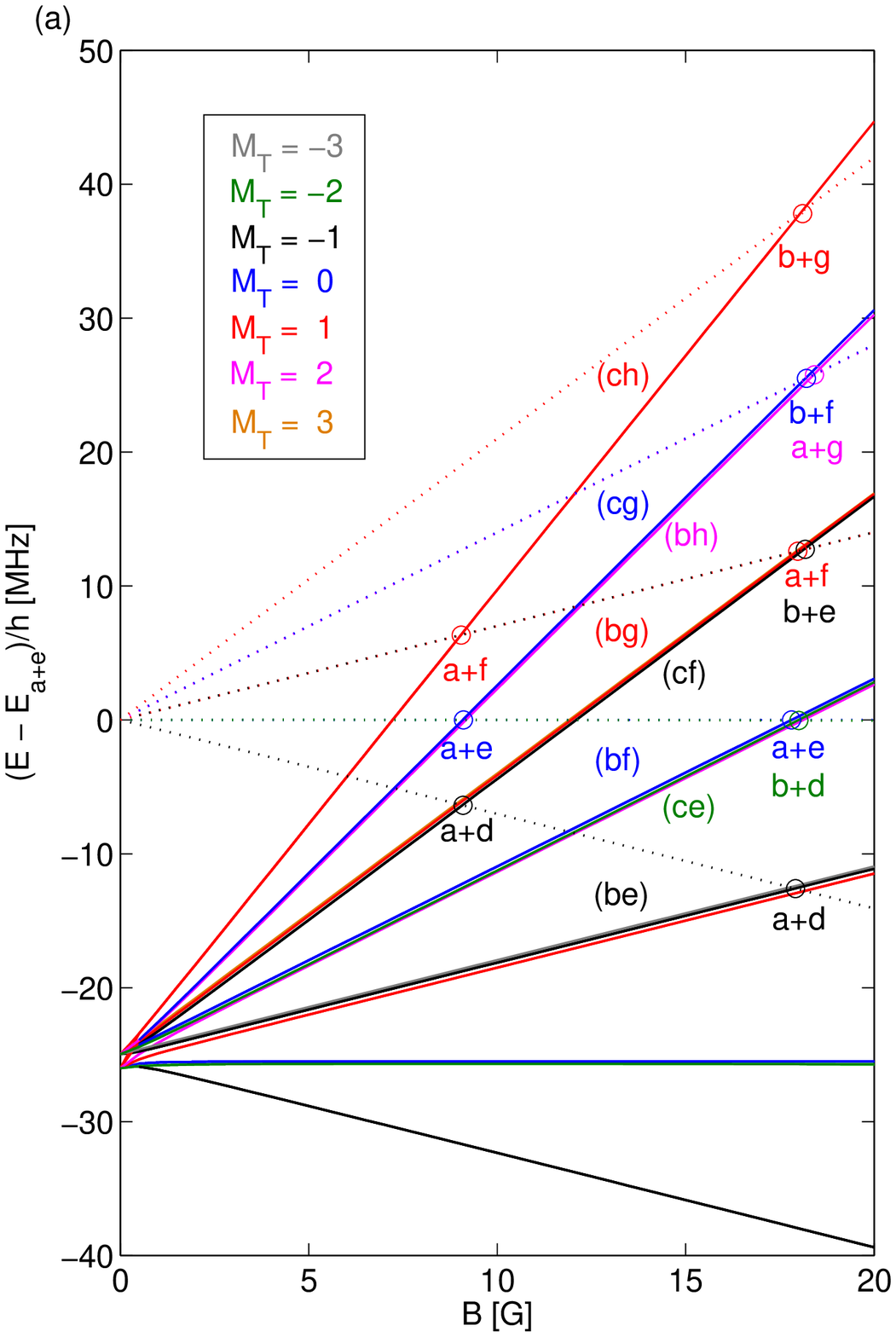}\
 \end{minipage}
\begin{minipage}[h]{0.49\columnwidth}
		\includegraphics[width=0.99\columnwidth, clip]{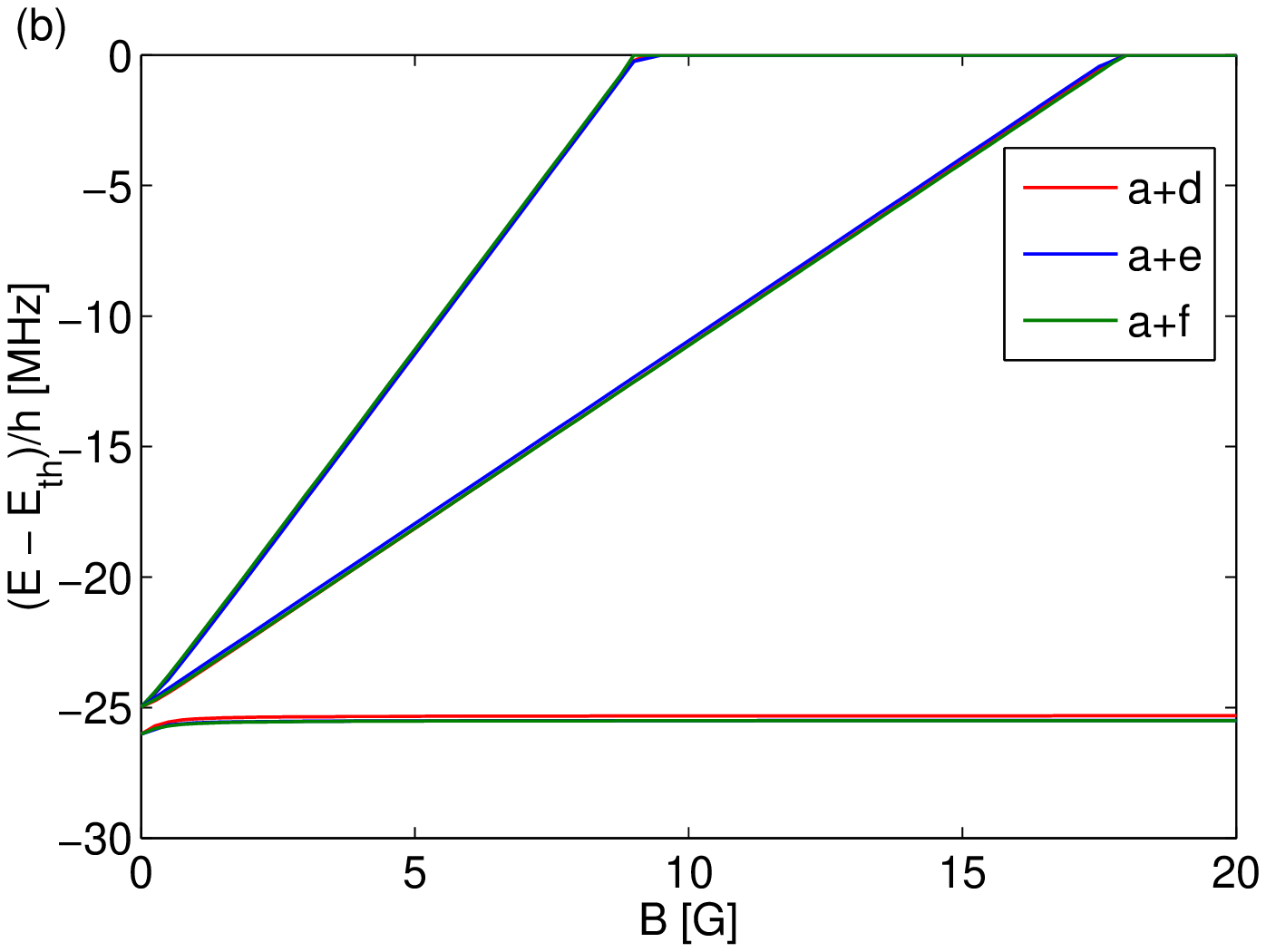}\\
		\includegraphics[width=0.99\columnwidth, clip]{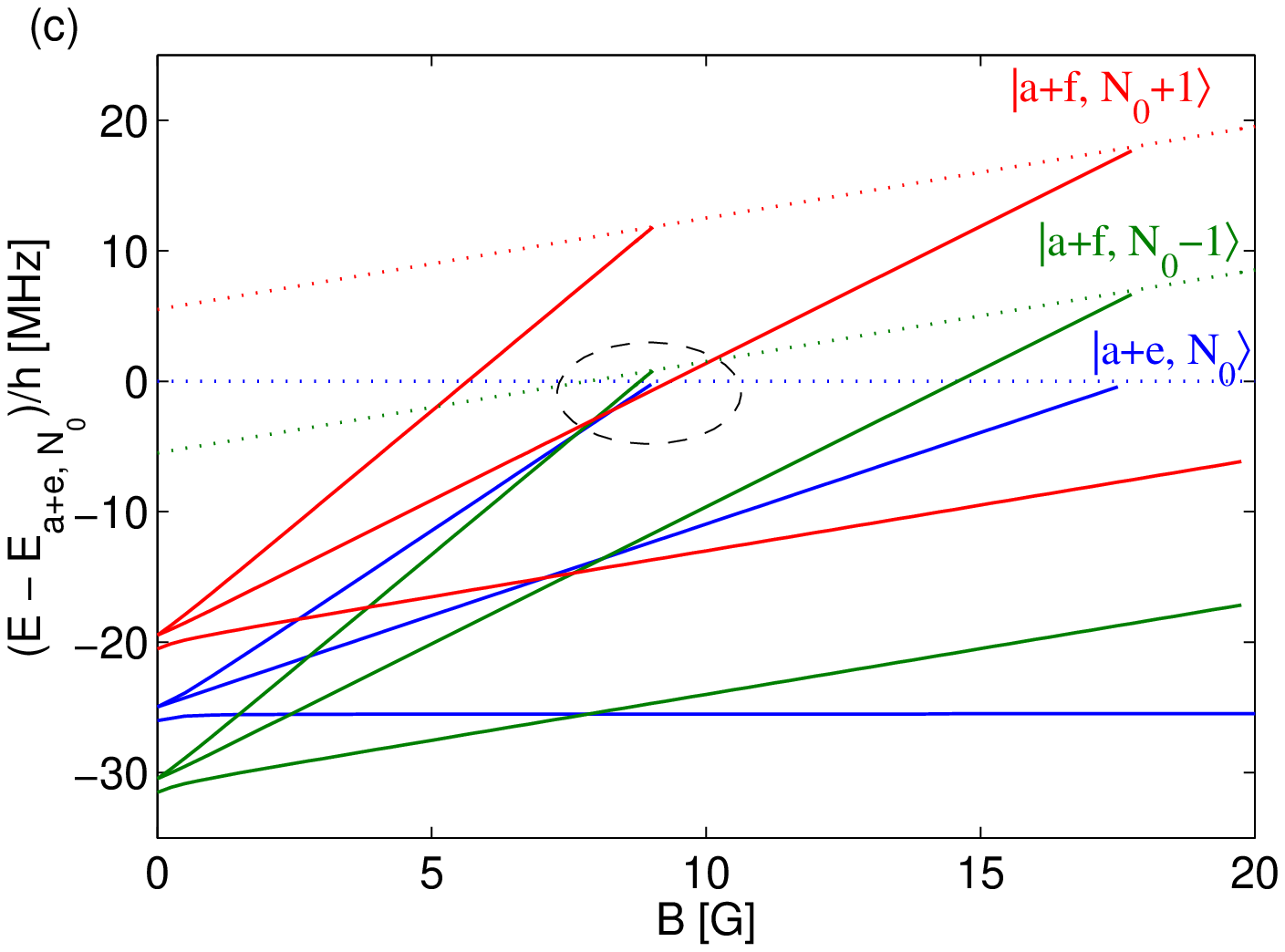}
		\end{minipage}
	\caption{Bound state structure and Feshbach resonances of \rbs. 
	Energies are shown as a function of magnetic field. 
	Bound states are shown as solid lines, and labelled with the channel $(\alpha\beta)$ in which they are concentrated. 
	Collision thresholds are shown with dotted lines. 
	In (a), energies are shown relative to the $a+e$ entrance channel threshold.
	Resonances occur when a bound state crosses a collision threshold of the same $M_T$. 
	These are shown with points and labelled with the entrance channel $\alpha + \beta$.
	Note that we only show the thresholds of channels containing a resonance.
	Colors indicate different $M_T$, as labelled.
	The Zeeman effect is close to linear in the field range shown, and each bound state is concentrated in a single channel.
	This gives rise to the grouping of resonances around 9\,G and 18\,G.
	In (b), the bound state spectra of the $a+d$, $a+e$ and $a+f$ channels are shown relative to their own thresholds. 
		In (c), the dressed channel energies are shown relative to that of the \ket{a+e, N_0} threshold.
	Here, the rf frequency is 5.5\,MHz.
	The bound state structure shown in (a) and (b) results in a number of bound states becoming degenerate with each other and the \ket{a+e, N_0} threshold near 9\,G.
	The bound state causing the $a+e$ resonance is coupled most strongly to states in the \ket{a+f, N_0 \pm 1} channels, which are indicated with the dashed circle in panel (c).
	}
	\label{fig:boundstates}
\end{figure}
%%%%%%%%%%
All collision channels of \rbs\ have a similar value of $a_\p{bg}$, because of the similarity of the singlet and triplet scattering lengths. 
Another consequence of this is that the highest vibrational bound states for all possible $F$ have approximately the same binding energy at $B = 0$.
Figure~\ref{fig:boundstates}a shows the $F = 1, 2,$ and $3$ bound states closest to the $(f_1 = 1) + (f_2 = 2)$ threshold.
At zero field, the $F = 1$ and $F=3$ bound states are degenerate, with the $F = 2$ state approximately $h\times 1$\,MHz deeper. 
The atomic and molecular states are split into their Zeeman components at nonzero field.
At magnetic fields above 1\,G, each bound state is strongly concentrated ($>90$\,$\%$) in a single channel  $\alpha + \beta$, and is labelled $(\alpha\beta)$.
The energy of each bound state then remains at a constant offset from the threshold energy of its corresponding channel.
Also, the Zeeman effect is close to linear within the range of magnetic field considered here.
Consequently, for a fixed magnetic field, each pair of adjacent channels or bound states (of the same $f_1$ and $f_2$) have approximately the same energy gap.
This is illustrated in \reffig{fig:boundstates}b, which shows the bound states of the $a+d$, $a+e$ and $a+f$ channels.
Here, energies are shown relative to the threshold of each channel.
The three spectra are nearly identical, highlighting the symmetries inherent in the Zeeman interaction.
The combinations of atomic states from which channels of each $M_T$ can be formed, and the bound state energies at $B = 0$, explain the existence of the three Feshbach resonances near 9\,G and the eight near 18\,G shown in \reffig{fig:boundstates}a~\cite{kaufman09, notebfbg}.

Another consequence of the Zeeman effect being predominantly linear is that several channels can be resonantly coupled by a single rf frequency, as shown by the avoided crossings in \reffig{fig:rfen_2b}.
It is also possible to couple several bound states, and use the magnetic field to tune these coupled states through a collision threshold.
We consider atoms colliding in the $\ket{a+e, N_0}$ entrance channel, and choose an rf frequency close to the atomic Zeeman splitting 
at 9\,G.
The resulting energies of the $\ket{a+f, N_0 +1}$ and $\ket{a+f, N_0 - 1}$ bound states relative to the \ket{a+e, N_0} threshold are shown in \reffig{fig:boundstates}c, for an rf frequency of $\omega = 2\pi \times 5.5$\,MHz. 
There is near-degeneracy between the bound states causing the 9\,G resonances in the $\ket{a+e, N_0}$ and $\ket{a+f, N_0 - 1}$ channels, and that causing the 18\,G \ket{a+f, N_0 + 1} resonance.
Consequently, a pair colliding in the entrance channel are linked by spin exchange to a set of strongly rf-coupled bound states.

\begin{figure}[tb]
	\centering
		\includegraphics[width=0.48\columnwidth, clip]{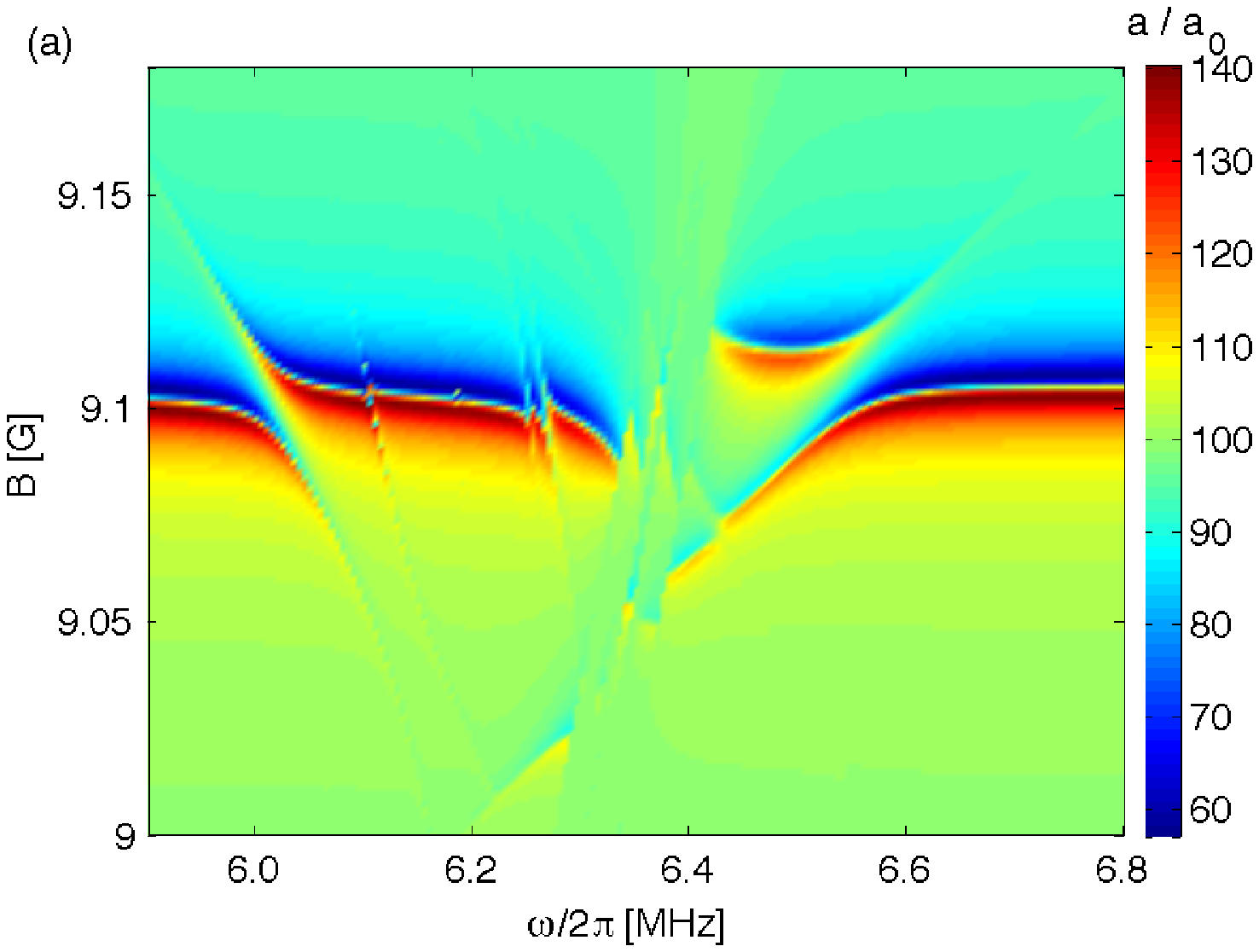}
\includegraphics[width=0.48\columnwidth, clip]{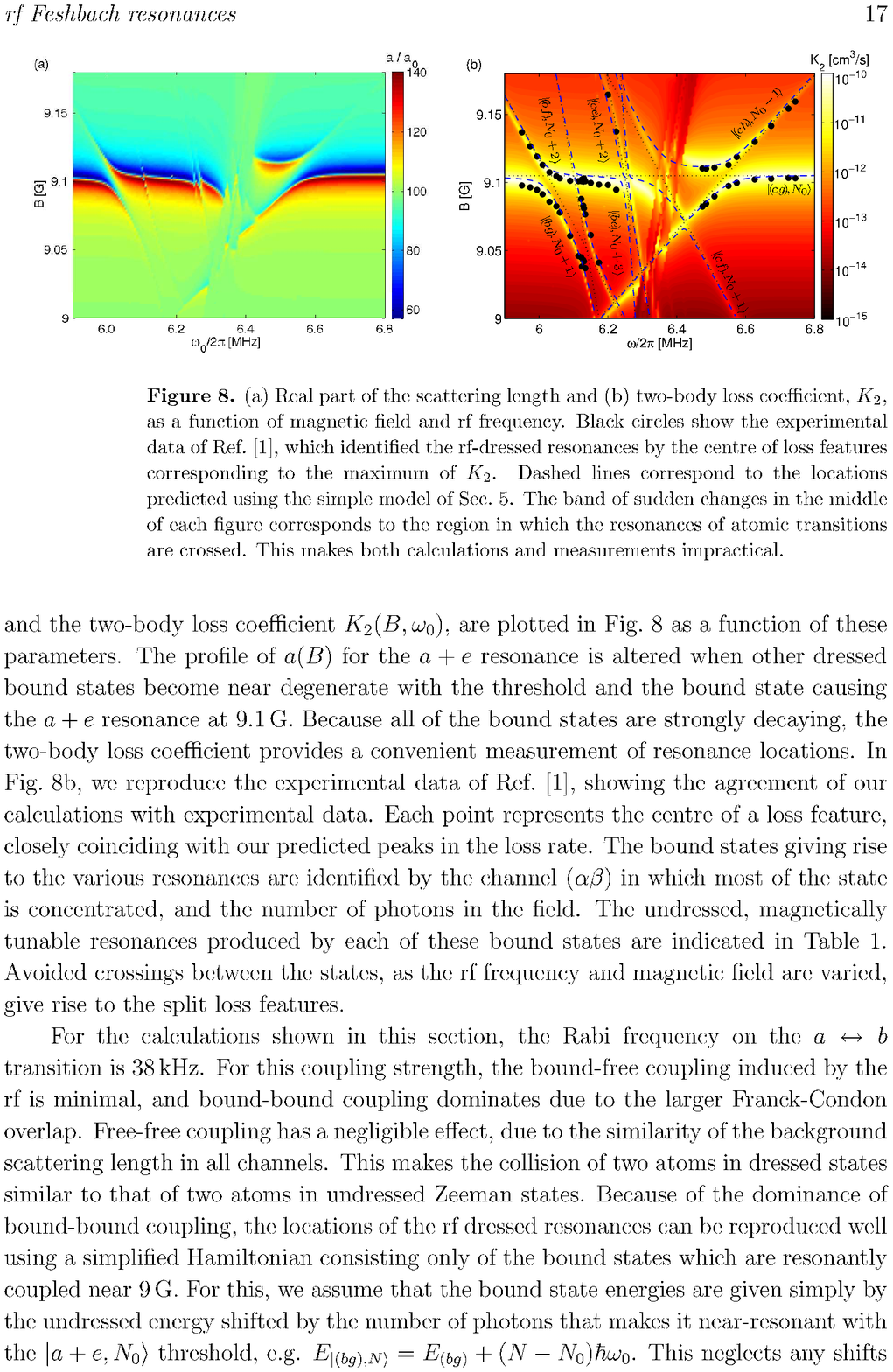}
	\caption{(a) Real part of the scattering length and (b) two-body loss coefficient, $K_2$, as a function of magnetic field and rf frequency. Black circles show the experimental data of Ref.~\cite{kaufman09}, which identified the rf-dressed resonances by the centre of loss features corresponding to the maximum of $K_2$. Dashed lines correspond to the locations predicted using the simple model of Sec.~\ref{sec:rb} which includes only bound states. The band of sudden changes in the middle of each figure corresponds to the region in which the resonances of atomic transitions are crossed. This makes both calculations and measurements impractical.}
	\label{fig:bw0_40khz}
\end{figure}

Several other bound states contribute to the rf-dressed scattering around 9\,G, leading to complicated variation of scattering properties with magnetic field and rf frequency.
These are not shown in \reffig{fig:boundstates}c for reasons of clarity.
The real part of the scattering length $a(B, \omega)$, and the two-body loss coefficient $K_2(B, \omega)$, are plotted in \reffig{fig:bw0_40khz}. 
The broad, horizontal band in both plots corresponds to the undressed $a+e$ resonance.
Coupling of the underlying bound state to others creates several secondary features at magnetic fields that depend on the rf frequency.
When this field is close to 9.1\,G, the $a+e$ resonance is split.
The rf-induced avoided crossing then changes the magnetic field at which each bound state crosses the entrance channel threshold.
Although it is therefore possible to suppress losses at the centre of a resonance or move its location, this will in general also suppress or move the resonant enhancement of the scattering length.
This is shown by the variation in scattering length of \reffig{fig:bw0_40khz}a, which follows the same trends as the losses in \reffig{fig:bw0_40khz}b.
Loss rates generally have a peak close to Feshbach resonances, and are therefore commonly used for deducing resonance locations.
In \reffig{fig:bw0_40khz}b, we reproduce the experimental data of Ref.~\cite{kaufman09}, showing its close agreement with our calculations.
Each point represents the centre of a loss feature.
The bound state giving rise to each feature is identified by the rf-dressed channel $\ket{(\alpha_1\alpha_2), N}$ in which it is concentrated.
The undressed, magnetically tunable resonances produced by each of these bound states are shown in \reffig{fig:boundstates}a.

% Coupled bound states
For the Rabi frequency of 38\,kHz used here, the bound-free coupling induced by the rf is minimal, and bound-bound coupling dominates due to the larger Franck-Condon overlap. 
Because of the dominance of bound-bound coupling, the locations of the rf dressed resonances can be reproduced well using a simplified Hamiltonian consisting only of the bound states which are resonantly coupled near 9\,G. 
For this, we assume that the bound state energies are given simply by the undressed energy shifted by the number of photons that makes it near-resonant with the \ket{a+e, N_0} threshold, 
e.g. $E_{|(bg),N\rangle} = E_{(bg)} + (N - N_0)\hbar\omega$.
This neglects any shifts in the threshold energies due to the intensity of the light.
We also use the atomic Rabi frequencies to represent the coupling between bound states, e.g.
$\Omega_{(bg)\leftrightarrow(cg)} = \Omega_{b \leftrightarrow c}$.
This neglects any differences in the molecular wavefunctions, the overlap of which should be close to unity for this case of narrow resonances.
We diagonalise the Hamiltonian and find the fields and rf frequencies for which the dressed bound states become degenerate with the $\ket{a+e,N}$ threshold. 
These are shown as dashed lines in \reffig{fig:bw0_40khz}. 
The good agreement of this simpler calculation confirms the dominance of bound-bound coupling in the effects of the rf on the scattering properties.

We note that unlike the losses found for \lisix\ in Sec.~\ref{sec:li}, which are into exit channels coupled by the rf, the losses here are due to the properties of the magnetically tunable resonances themselves. 
All of these resonances are closed-channel dominated and strongly decaying, due to inelastic spin relaxation loss into lower channels of the same $M_T$~\cite{kaufman09}. 
Each has a width of order mG and a resonance length of $ a_\p{res} \lesssim a_\p{bg}/2$. 
Consequently, the strongest losses in \reffig{fig:bw0_40khz} are independent of Rabi frequency, and a secondary feature due to a bound state from a channel with $N$ photons grows in strength proportional to $\Omega^{2|N - N_0|}$.

%%%%%%%%%%%%%%%%%%%%%%%%%%%%%%%%%%%%%%%
\subsection{{\lisix}: free-free coupling}
\label{sec:liff}

\begin{figure}[tb]
	\centering
		\includegraphics[width=0.6\columnwidth, clip]{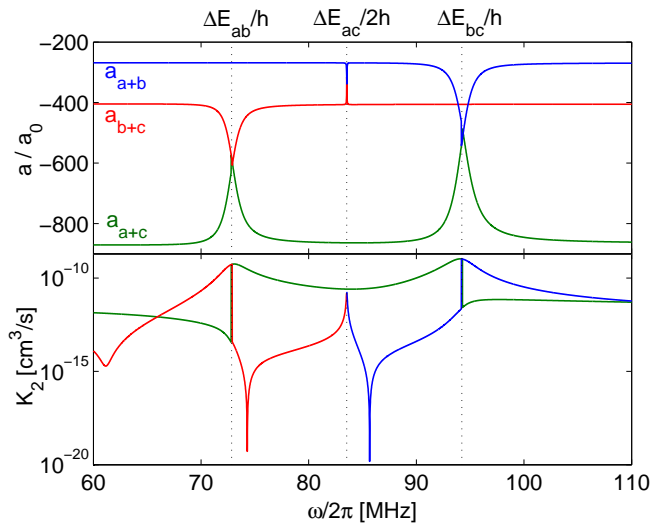}
	\caption{Scattering length and loss rate as a function of rf frequency $\omega/2\pi$. 
	The magnetic field is 250\,G, and the $\sigma^{-}$ rf has a Rabi frequency of 1\,MHz.
	Significant changes in scattering properties occur close to the resonant frequencies of the $a \leftrightarrow b$ and $b \leftrightarrow c$ atomic transitions. 
	A two-photon $a \leftrightarrow c$ transition is also driven.
	The loss properties of each channel change dramatically as the transition frequencies are crossed. 
	Away from the transition frequencies, small changes in scattering length are achievable with low loss rates.
	}
	\label{fig:liff}
\end{figure}
It is possible to control scattering properties by creating rf-dressed atomic states. 
This is done routinely in the creation of clock states~\cite{audoin}.
For rf of a frequency close to an atomic transition, the dressed atoms have a substantial admixture in the two coupled Zeeman states.
A collision between two dressed atoms then samples the scattering properties of all the corresponding undressed two-body channels.
For such free-free coupling to have an effect requires the scattering length in each undressed channel to be substantially different. 
In \rbs, for example, all channels have very similar background scattering lengths.
This makes the collision of two atoms in dressed states similar to that of two atoms in the absence of rf.
By contrast, in \reffig{fig:liff} we show the example of \lisix\ scattering.
The broad, overlapping Feshbach resonances in the $a+b$, $a+c$ and $b+c$ channels make the difference in scattering lengths between the channels have a nontrivial variation with magnetic field.
Large changes in scattering length are produced for rf frequencies close to the atomic transitions, over a range comparable to the Rabi frequency.

The scheme described in this subsection does not create a Feshbach resonance.
It does, however, sample the scattering lengths in different Zeeman channels, including any Feshbach resonances those channels support.
Very close to the atomic transition frequencies, losses will be a problem for this scheme, as shown in the lower panel of \reffig{fig:liff}. 
However, such a method could be useful for creating small changes in the scattering length, and may remain an option in the absence of bound states suitable for using the methods of the previous two subsections.
In the example given here, a 5\% change in scattering length can be achieved with loss rates of order $10^{-13}$ cm$^{3}/$s.

\section{Conclusions}
\label{sec:conclusion}

We have developed a technique for studying the collisions of cold atoms in the presence of radio-frequency radiation. 
Building on the three-parameter model of Feshbach resonances presented in Ref.~\cite{hanna09}, we have incorporated a frame transformation to a basis in which Zeeman states are coupled together by rf radiation.
Our studies have shown that rf provides useful control capabilities when used in conjunction with one or more Feshbach resonances - either by coupling together bound states with which the colliding atoms interact~\cite{kaufman09}, or by directly coupling the colliding pair to a molecular state.
For the latter case, we have found that the use of a halo molecule reduces the rf power required for control.
Also, some ranges of rf frequency and magnetic field exist for which a bound state can be coupled without causing losses.
This requires the use of circularly polarised rf.

The accuracy and speed with which rf can be controlled, and the potential of tuning several scattering lengths of a multicomponent gas, give our work relevance to many current experimental programs in atomic and molecular collisions.
Other candidates for such an enhancement of control possibilities include $^{40}$K, which has broad resonances in the $a+b$ and $a+c$ channels. 
Another example more like the \rbs\ calculations of Sec.~\ref{sec:rb} is $^{6}$Li-$^{40}$K, which has several narrow resonances around 160\,G. 
The $a+a$ channel could allow bound-bound coupling without the strong decay inherent to the excited state \rbs\ resonances.

\section{Acknowledgements}
P.S.J acknowledges partial support by the U.S. Office of Naval Research. 
We are grateful to Adam M. Kaufman, Russell P. Anderson and David S. Hall for allowing us to reproduce the experimental data from our previously published collaboration~\cite{kaufman09}, and thank them as well as Timur Tscherbul for many fruitful and stimulating discussions.

\section*{References}

\bibliographystyle{unsrt}

\bibliography{rfrefs}

\begin{thebibliography}{10}
\bibitem{notebfbg}
In addition to the six 18\,G resonances measured in Ref.~\cite{kaufman09}, our
  calculations show two very narrow resonances in the $b+f$ and $b+g$ channels
  at 18.2\,G and 18.1\,G, respectively.
\end{thebibliography}

\end{document}